\documentclass[twocolumn,showpacs,amsmath,amssymb]{revtex4}

\usepackage{graphicx}
\usepackage{dcolumn}
\usepackage{bm}

\begin{document}
\title{A new fast reconnection model in a collisionless regime}
\author{David Tsiklauri}

\affiliation{Joule Physics Laboratory,
Institute for Materials Research, University of Salford, Manchester, M5 4WT,
United Kingdom}
\date{\today}

\begin{abstract}
Based on the first principles 
(i.e. (i) by balancing the magnetic field advection 
with the term containing electron pressure tensor
non-gyrotropic components in the generalised Ohm's law;
(ii) using the conservation of mass; and (iii) assuming that the weak magnetic field 
region width, where
electron meandering motion supports electron pressure tensor
off-diagonal (non-gyrotropic) components, 
is of the order of electron Larmor radius;)
a simple model of magnetic reconnection in a collisionless
regime is formulated. 
The model is general, resembling its 
collisional Sweet-Parker analogue in that 
it is not specific to any initial configuration e.g. Harris type tearing unstable current
sheet,
X-point collapse or otherwise.
In addition to its importance from the 
fundamental point of view, the collisionless reconnection model 
offers  a much faster  reconnection rate ($M_{c'less}=( c / \omega_{pe})^2 / (r_{L,e} L)$)
than Sweet-Parker's classical one ($M_{sp}=S^{-1/2}$). 
The width of the diffusion region (current sheet) in the
collisionless regime is found to be $\delta_{c'less}=( c / \omega_{pe})^2 /r_{L,e}$, which
is independent of global reconnection scale $L$ and is only prescribed by micro-physics
(electron inertial length, $c / \omega_{pe}$, and electron  Larmor radius, $r_{L,e}$).
Amongst other issues, the fastness of the reconnection rate alleviates e.g. 
the problem of interpretation of solar flares by means of reconnection, 
as for the typical
solar coronal parameters the obtained 
collisionless reconnection time can be a few minutes,
as opposed to Sweet-Parker's equivalent value of $<$ a day.
The new theoretical reconnection rate is compared to the
MRX (Magnetic Reconnection Experiment) device experimental data by \citet{yamada,ji} and a good 
agreement is obtained.
\end{abstract}	

\pacs{52.35.Vd; 96.60.Iv; 52.65.Rr; 45.50.Dd; 96.60.pf; 96.60.qe}

\maketitle
\section{Motivation}
Magnetic reconnection is the process that enables to convert magnetic energy
into heat and kinetic energy of accelerated particles \cite{pf00,biskamp,bp07}. 
In many astrophysical and laboratory plasma
situations plasma beta (the ratio of thermal and magnetic pressures) 
is much smaller than unity. Thus,
magnetic reconnection has attracted a considerable attention 
as the plasma heating and charged particle acceleration mechanism. 
It is believed that magnetic reconnection
is responsible for solar \cite{1997Natur.386..811I,2008AdSpR..42..895W} and stellar flares 
\cite{2008A&A...481..799S,2008ApJ...676L..69C}. 
It has been also observed in the
Earth geomagnetic tail \cite{2008JGRA..11307S27R}.

This work has several motivations:

(i) The longstanding problems with resistive MHD (Magnetohydrodynamic), i.e. 
{\it collisional}
description of the magnetic reconnection has been 
its too slow rate if a consistent, the-first-principles approach
is used  \cite{sweet,parker}; or a fast rate \cite{petschek}, but a lack of a 
proper motivation (\cite{biskamp}, p. 79) (one can still get the fast reconnection if the
resistivity is enhanced in the diffusion region, as one expects in
physical applications). In the {\it collisionless}
regime there is no known simple, analytical reconnection rate (e.g. analogous to
Sweet-Parker, or Petschek rates). What does exist, however, 
is a bulk of mostly
numerical simulation 
work \cite{hesse99,birn01,pritchett01,th07,th08}.
It should be mentioned that a significant progress (including analytical) 
has been made
in study of the details of collisionless reconnection such as the diffusion
region structure, outflows, reconnection rates \cite{kuz98,hesse99,birn07,klimas08}.
However these were either specific to a narrow class of
collisionless reconnection models, namely tearing unstable 
Harris current sheet; or too slow reconnection rates were obtained \cite{vp95}.
Some numerical simulation models 
\cite{kuz98,hesse99,birn01,pritchett01,birn07,th07,th08,klimas08} can 
produce as fast reconnection rates
as $M>0.5$ where $M$ is the inflow Alfv\'enic Mach number.
The lack of a simple, analytical model
of collisionless reconnection ({\it that is not specific to any initial configuration}) 
has been somewhat hampering the progress in understanding. 

(ii) Quite often, particularly in space plasmas, there is a good reason to
go beyond resistive, single-fluid MHD. A simple line of
reasoning can be outlined on an example of e.g. solar corona.
Fixing coronal temperature at $1.0 \times 10^6$ K, Coulomb logarithm at 18.1, the 
Lundquist number (using Spitzer resistivity) is 
$3.7 \times 10^{11}$. Here $L=10^5$ m was used.
One of the arguments for going beyond resistive MHD is comparing
typical width of a Sweet-Parker current 
sheet $\delta_{sp}=S^{-1/2} L=0.16$ m to the ion inertial length.
Typical scale associated with the Hall term in the generalised Ohm's law at which deviation from
electron-ion coupled dynamics is observed is, $c/\omega_{pi} =  7.2$ m. Here 
particle density of $n=1.0 \times 10^{15}$ m$^{-3}$ is used.
Hence, the fact that $ c/(\omega_{pi} \delta_{sp})=44 \gg 1 $ warrants
going {\it beyond single fluid resistive MHD approximation} (a 
similar conclusion is reached by \citet{yamada} in their Figure 12 based on laboratory plasma 
experiment known as MRX (Magnetic Reconnection Experiment)).

(iii) Previous results on collisionless reconnection both in tearing unstable 
Harris current sheet \cite{kuz98,hesse99,birn01,pritchett01} and stressed X-point collapse 
\cite{th07,th08} has shown that magnetic field is frozen into electron fluid
and the term in the generalised Ohm's law 
that is responsible for breaking the frozen-in condition is
electron pressure tensor off-diagonal (non-gyrotropic) components. 
Thus, a need for inclusion of the
{\it electron pressure tensor non-gyrotropic components} in a 
model of collisionless 
reconnection has become clear.

(iv) There is a growing amount of work \cite{bp07,th07,th08} that 
suggests that in the collisionless regime,
on the scales less than $c/\omega_{pi}$ magnetic field is frozen into the 
electron fluid rather than bulk of plasma.
One can write in general $\vec V_e=\vec V_i- \vec j / (e n)$. This relation clearly
shows that in collisional regime (when the number density $n$ is large), the
difference between electron and ions speeds diminishes $V_e=V_i=V$. However, as one enters
collisionless regime (when the number density $n$ is small) the deviation between
electron and ion speeds starts to show.
In \citet{th08}
we  proposed a possible explanation why the reconnection is fast when
the Hall term is included. Inclusion of the latter means that
in the reconnection inflow magnetic field is frozen into {\it electron} fluid.
As it was previously shown in \citet{th07} (see their Figs.(7) and (11))
speed of electrons, during the reconnection peak time, is
 at least 4-5 times greater than that of ions. This means that electrons can
 bring in / take out the magnetic field attached to them into / away from 
 the diffusion region
 much faster than in the case of single fluid MHD which 
 does not distinguish between
 electron-ion dynamics. Thus, a need for inclusion {\it magnetic field transport by electrons}
  in a model of collisionless reconnection has become clear.
 
(v) A pioneering work of \citet{yamada} has 
demonstrated a clear transition from collisional to collisionless
reconnection regimes by varying the plasma density, and established that
in the collisionless regime the reconnection rate is much greater than the
Sweet-Parker rate. Despite a well motivated explanation of
their experimental results, one was surprised to see only
experimental data points and numerical simulation results on their Figure 12.
Thus, we set out with a task
of formulating a simple model of magnetic reconnection in a collisionless
regime. In what follows we shall be using
well-motivated arguments of Sweet-Parker model, but shall be applying it
to the collisionless case.

\section{The Model}

It is instructive to re-iterate key points of the Sweet-Parker model.
The first step towards the derivation of the reconnection rate is that the
plasma outflow speed from the diffusion region is of the order of the Alfv\'en
speed, $V_{ outflow}=V_A=B_0/\sqrt{\mu_0 m_i n}=B_0/\sqrt{\mu_0 \rho}$. This follows from the
following consideration: taking the fluid to be incompressible
and assuming a steady state condition, one obtains \cite{golub}, p. 285
\begin{equation}
\frac{\rho V_{ outflow}^2}{2}=p_i-p_o,   
\end{equation}
where $p_i$ and $p_o$ are thermal pressures inside and outside the diffusion region.
This pressure difference can be set to the magnetic pressure $B_0^2/(2 \mu_0)$.
From which the above result $V_{ outflow}=V_A$ readily follows.

The second step is applying the continuity equation
\begin{equation}
V_{ inflow} L =V_A \delta, 
\end{equation}
where $L$ and $\delta$ are the diffusion region length and width respectively.

The third step is using the generalised Ohm's law
(e.g.  \cite{bp07} p. 108)
\begin{equation}
  \vec{E} = - \vec{v}_e \times \vec{B} 
            - \frac{\nabla \cdot \vec{P}_e}{n_e e}
            - \frac{m_e}{e} \left(\frac{\partial \vec{v}_e}{\partial t} 
            + \left( \vec{v}_e \cdot \nabla \right) \vec{v}_e \right) + \eta
	    \vec j,   
\end{equation}
where $\vec{E}$ and $\vec{B}$ are electric and magnetic fields, $\vec{v}$ is plasma 
 velocity, $\vec{P}$ is pressure tensor ($3 \times 3$ matrix), $n$ is 
plasma number density, $m$ is  mass and $e$ is electric charge.
Subscript $e$ stands for an electron.
In Eq.(3) we balance two terms advection ($\vec{v}_e \times \vec{B}$) and
resistive diffusion ($\eta \vec j$):
\begin{equation}
V_{ inflow} B_0 = \eta j =\eta B_0/(\mu_0 \delta),  
\end{equation}
where $\vec j = (\nabla \times \vec B)/ \mu_0$ is used, with $\nabla \approx 1 / \delta$.
Note that in the (collisional) resistive MHD regime all other terms
in the generalised Ohm's law are insignificant and $V_e=V_i=V$.
Finally, substituting $(1 / \delta)$ from Eq.(2) and after some simple algebra
one arrives at 
\begin{equation}
M_{sp} \equiv V_{ inflow}/V_A=S^{-1/2}, 
\end{equation} Where $S=\mu_0 L V_A/ \eta$,
the Lundquist number has been introduced. Eq.(5) constitutes 
the classical Sweet-Parker reconnection
rate. 

In order to obtain collisionless reconnection rate in an analogy with the
above derivation of the classical Sweet-Parker rate,
we now balance the advection ($\vec{v}_e \times \vec{B}$) with 
electron pressure tensor (${\nabla \cdot \vec{P}_e}/({n_e e})$), because 
now Spitzer resistivity in the collisionless regime is not large enough and also
$V_e \not = V_i=V$.
Previous results of  the reconnection in the collisionless
regime, both in tearing unstable 
Harris current sheet \cite{hesse99,birn01,pritchett01} and stressed X-point collapse 
\cite{th07,th08}, have shown that: (i) the term in the generalised Ohm's law 
that is responsible for breaking the frozen-in condition is
 electron pressure tensor non-gyrotropic components
 and (ii) the magnetic field is frozen into electron fluid.
 \citet{kuz98}  
 formulate two dimensional model,  which shows that
 in a steady state, at the magnetic null (X-point)
 where the electron flow velocities and magnetic field are zero,
 the only terms in the generalized Ohm's law that can support
 out-of-plane electric field (and hence the reconnection) are 
 electron pressure tensor non-gyrotropic components (see their Eqs.(4)-(6)):
 \begin{equation}
 E_y^{NG}= - \frac{1}{n e}\left(\frac{\partial P_{xy}}{\partial x} + 
 \frac{\partial P_{zy}}{\partial z}\right). 
 \end{equation}
Here the two $x$-  and $z$-coordinates are along (outflows) and across (inflows) 
the current sheet
respectively. 
Further simplification can be made observing that
$P_{xy} \approx P_{zy}$ but $\partial / \partial z \gg \partial / \partial x$
because width of the current sheet is much smaller than its length:
\begin{equation}
 E_y^{NG} \approx - \frac{1}{n e}
 \frac{ P_{zy}}{ \delta_{c'less}}, 
 \end{equation}   
 where $\delta_{c'less}$ is the width of the current sheet in the
 collisionless regime ($\partial / \partial z \approx \delta_{c'less}$).
Thus in an analogy with Eq.(4) we now have 
\begin{equation}
 E_y^{NG} = V_{ e,inflow} B_0  \approx - \frac{1}{n e}
 \frac{ P_{zy}}{ \delta_{c'less}}. 
 \end{equation}
 Note that the latter equation is similar to Eq.(11) from \citet{hesse99}.
In addition to balancing the different terms, note also
that the two crucial  differences from Eq.(4) are that (i) we use electron speed 
$V_{e,inflow}$ because magnetic field advection on scales less than 
$c/\omega_{pi}$ is done by electrons and (ii) width of the diffusion region
(current sheet) is $\delta_{ c'less}$ which we shall specify below.

Let us now specify electron pressure tensor non-gyrotropic component
$P_{zy}$. Based in earlier works, which use second moment of 
Vlasov equation (i.e. multiplying the Vlasov equation by $v_i v_j$
and integrating over the velocity space), \citet{kuz98}
give explicit expression for the  electron pressure tensor components.
Intermediate expression for $P_{zy}$ in the static case, which is
generic and yet is not specific to tearing unstable 
Harris current sheet configuration, is given by Eq.(14) in \citet{kuz98}:
\begin{equation}
P_{zy}= - \frac{P_{zz}}{2 \Omega_x} \frac{\partial v_{ez}}{\partial z},
\end{equation}
where $\Omega_x=(eB_0/m_e)(z / \delta_{ c'less})$ takes into account
linear variation of magnetic field with distance in the vicinity of the
magnetic null. The electron pressure tensor non-gyrotropic (off-diagonal) components
are generally much smaller than gyrotropic (diagonal) ones.
The deviations from gyrotropic pressure are possible due to
electron meandering motion in the regions of weak magnetic field
close to the x-point \cite{hs97}. Away from the x-point 
the particles are magnetized and pressure is isotropic.
Thus, it would be reasonable to assume that electron meandering
motion will be effective up to $z \approx r_{L,e}$.
Here  $r_{L,e}=v_{th,e} / \omega_{ce}$ is the electron Larmor radius.
$v_{th,e}=\sqrt{k_B T / m_e}$ is electron thermal speed and 
$\omega_{ce}=e B /m_e$ is electron cyclotron frequency. 
Further away from the magnetic null, when $z \gg r_{L,e}$
then pressure becomes isotropic. Thus in the above expression for $\Omega_x$
we set $z \approx r_{L,e}$, i.e. $\Omega_x=(e B/m_e)(r_{L,e} / \delta_{ c'less})$.
In Eq.(9) we can also assume the  gyrotropic pressure component $P_{zz}$
is of the order of the magnetic pressure $B_0^2/(2 \mu_0)$, set  $v_{ez}$
as the electron inflow speed $V_{ e,inflow}$, also estimate 
$\partial / {\partial z} \approx 1 / \delta_{c'less}$.
Thus for the $P_{zy}$ we have
\begin{equation}
P_{zy}= -\frac{B_0^2}{2 \mu_0} \frac{m_e}{e B_0} \frac{\delta_{c'less}}{r_{L,e}} 
\frac{V_{ e,inflow}}{\delta_{c'less}}.
\end{equation}

Naturally, the next step is to specify width of the diffusion region
in the collisionless regime, which we do by using conservation of
mass, but again taking into account that
now magnetic field is advected by the electrons. Thus, in lieu of Eq.(2) we have
\begin{equation}
V_{ e,inflow} L =V_{e,A} \delta_{ c'less},    
\end{equation}
where $V_{e,A} = B_0 / \sqrt{\mu_0 n m_e} = V_A \sqrt{m_i/m_e}$ is the {\it electron} Alfv\'en speed.

Thus substituting Eq.(10) into Eq.(8), using Eq.(11) and defining
the collisionless inflow Alfv\'enic Mach number, $M_{ c'less}$ as 
$M_{ c'less}=V_{ e,inflow}/V_{e,A}$,
 after simple algebra (also noting that $n e^2/m_e = \omega_{pe}^2 \epsilon_0$ and 
 $c=1/\sqrt{\mu_0 \epsilon_0}$) 
 we obtain
\begin{equation}
M_{ c'less} \equiv \left(\frac{c}{\omega_{pe}}\right)^2 \frac{1}{r_{L,e} L}.
\end{equation}
Note that factor $1/2$ has been omitted because this rate is a crude estimate.

Eq.(12) can be regarded as the main result of this work, as it provides
the collisionless reconnection rate independent of an initial configuration.

Simple analysis shows that collisionless reconnection rate $M_{ c'less}=
1.3 \times 10^{-4}$
e.g. for solar coronal parameters 
($n=1.0\times 10^{15}$ m$^{-3}$, $T=1.0 \times 10^6$ K, $L=10^5$ m, 
$B=0.01$T (100 Gauss and hence $V_A=6.9 \times 10^6$ 
m s$^{-1}$), $S=3.7 \times 10^{11}$)
is two orders of magnitude faster
than the classical Sweet-Parker rate $M_{sp}=1.6
\times 10^{-6}$. 
In the context of solar flares this means that the collisionless
model, presented here, effectively alleviates longstanding problem
of the interpretation of solar flares by means of magnetic reconnection.
The reconnection rate is also interpreted as the ratio of Alfv\'en
time ($\tau_A = L / V_A \approx 0.0145$ s) and
resistive (or reconnection) times. This means in the Sweet-Parker
model resistive (or reconnection) time is $0.0145 / S^{-1/2}$ s $= 0.1$ days.
While our model provides reconnection time of $0.0145/(1.3 \times 10^{-4})$ s $= 2$ minutes,
which is commensurate of solar flare observations. It should be mentioned
that Petschek model also can provide an appropriately fast
reconnection rate, although its justification from the
fundamental point of view is not without a debate \cite{biskamp}.

We now shall compare the master equation Eq.(12) to the MRX experimental data
of \citet{yamada}. Their figure 12 presents how the empirical
reconnection rate scales with the parameter $c/(\omega_{pi} \delta_{sp})$,
which is essentially the ratio of ion inertial length and 
the width of the Sweet-Parker current sheet. \citet{yamada} have clearly
established that in the collisionless regime,
i.e. when $c/(\omega_{pi} \delta_{sp}) \gg 1$ 
the normalised reconnection rate (after starting from the  
Sweet-Parker rate) attains values much larger than unity.
In order to compare our scaling law, Eq.(12) to the data from Ref.\cite{yamada},
we need to normalise the collisionless reconnection rate 
by the Sweet-Parker rate $S^{-1/2}$. Also, in order for the
results to be applicable in the collisional regime too, we combine
collisional and collisionless rates as follows:
$M=M_{sp}+M_{ c'less}$.
 Noting that 
$L=\delta_{sp} S^{1/2}$ we obtain,
\begin{equation}
M/M_{sp} = 1+ \frac{m_e}{m_i}\frac{\delta_{sp}}{r_{L,e}} \left(
\frac{c}{\omega_{pi} \delta_{sp} } \right)^{2}.
\end{equation}
Note that despite a factor $m_e/m_i$ appearing in Eq.(13), $M/M_{sp}$
does not depend on ion mass as $1/ \omega_{pi}^2 \propto m_i$. Also the same is
evident from the fact that $M_{ c'less}$ (according to Eq.(12)) only contains physical quantities
pertaining to an electron, and $M_{sp}$ is independent of ion mass, so should be their
sum. Our collisionless reconnection rate seems to fit the experimental data \cite{yamada} 
and the two-fluid 
simulation results \cite{breslau} well
in the range $c/(\omega_{pi} \delta_{sp}) \gg 1$.
It should be noted that each experimental data point
was obtained for different plasma parameters. Hence
each data point has its own $S$ associated with it.
$S$ changes from roughly $200$ to $800$ as we move in the 
lowest measured values of $c/(\omega_{pi} \delta_{sp}) =1.5$ to 
$c/(\omega_{pi} \delta_{sp}) =10$ in the horizontal axis (M. Yamada, 
private communication).
In order to provide theoretical {\it curves} rather than sets of the theoretical 
points
we argue that in Eq.(13) the dependence on $S$ is weak (via $\propto \delta_{sp}= L S^{-1/2}$).
Therefore, solid curve in Fig.1 is
produced using a fixed value of $S=200$. 
Other parameters used in Fig.~1 were $T=5$ eV, $L=0.4$ m
and $B=0.05$ Tesla.
Some deviation in the region of $c/(\omega_{pi} \delta_{sp}) \approx 1$
can be explained by, perhaps, several factors that were not taken into account
in our simple model. Let us mention the obvious two:
(i) Strictly speaking our reconnection rate Eqs.(12) or (13) was obtained
assuming a steady state, while as \citet{yamada} mention, the pull magnetic reconnection lasts
for about 40 $\mu$s. Although this is 
much longer than the typical Alfv\'en time $\leq \mu$s, 
yet one can argue that {\it nearly} steady state was achieved in the experiment.
Thus the comparison between the theory and experiment should be taken with caution.
(ii) Our theoretical reconnection rate, naturally, does not include contributions
from turbulence. Despite these shortcomings the match between the theory and experiment seems
good.

\begin{figure}
\includegraphics[scale = 0.7]{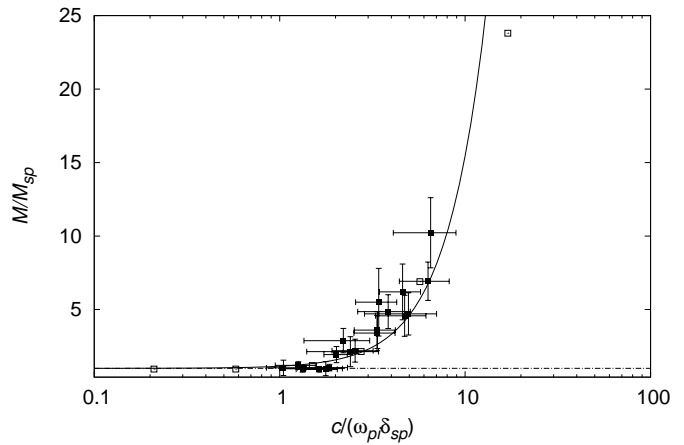}
\caption{\label{f1}
Solid curve shows
collisionless reconnection rate normalised to the Sweet-Parker rate according to Eq.(13) 
versus $c/(\omega_{pi} \delta_{sp})$.
Note that Hydrogen, Deuterium and Helium plasmas are all represented by the
same solid curve, as $M/M_{sp}$ is independent of an ion mass.
Data points with error bars correspond to MRX data \citet{yamada}. Note that in data we do not
distinguish between the species. Open square symbols are
the 2D two-fluid simulation results by \citet{breslau}.
Dash-dotted line represents the Sweet-Parker rate.}
\end{figure}

Based on Eqs.(11) and (12) it is also straightforward to derive
the width of the diffusion region (current sheet) in the
collisionless regime: 
\begin{equation}
\delta_{c'less}=M_{c'less} L=\left(\frac{c}{\omega_{pe}}\right)^2 \frac{1}{r_{L,e}}, 
\end{equation} 
which is independent of global reconnection scale $L$ and is only prescribed by micro-physics
(electron inertial length, $c/\omega_{pe}$, and electron  Larmor radius, $r_{L,e}$).

\citet{ji} have provided MRX laboratory experimental data and
2D Particle-In-Cell simulation results of scaling of the
electron diffusion region width, $\delta_e$, with the electron inertial length,
$c/ \omega_{pe}$. It was found that the experimental data can be
fitted with a straight line $ 8 c/ \omega_{pe}$, while PIC simulation data
consistently produced a less steep scaling of $ 1.6 c/ \omega_{pe}$.
Naturally, we tried to apply the diffusion region (current sheet)
width formula from our model (Eq.(14)) to  \citet{ji}.
Useful sketch to aid visualisation can be found in \citet{bp07}, p.91, Fig.3.1.
When plasma inflows in the diffusion region ions 
decouple from electrons at ion inertial length scale of $\approx c/ \omega_{pi}$,
while electrons deflect from the diffusion region (and hence
forming it)  at electron inertial length scale of $\approx c/ \omega_{pe}$.
One of the convincing findings of \citet{ji} was that 
the electron diffusion region width, $\delta_e$,
is independent of which ion species are used  in the experiment (H, D$_2$ or He).
Thus one can conclude that they indeed were observing electron diffusion
region. 
This corroborates the fact that 
at the electron inertial length scales
the magnetic field transport 
is done my electrons, as well as the currents are carried by electrons.
Our width of the diffusion region 
according to Eq.(14) is shown as solid curve in Fig.2.
A reasonably good fit is obtained, supporting the idea that Eq.(14) indeed
provides a good theoretical expression for the electron diffusion region width.

\begin{figure}
\includegraphics[scale = 0.7]{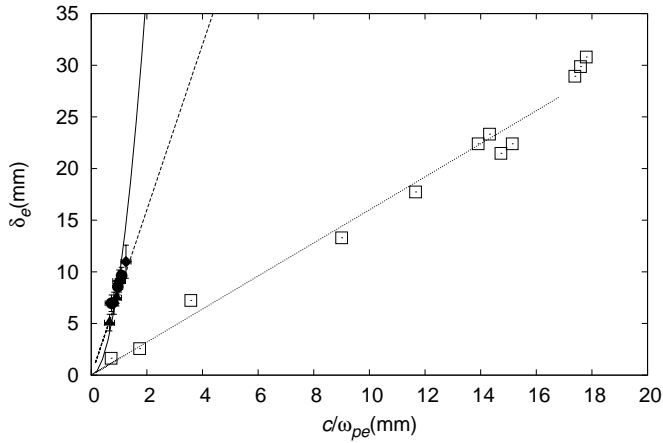}
\caption{\label{f1}
Solid  curve shows $\delta_{c'less}$ according Eq.(14).
Dashed and dotted lines represent
$ 8 c/ \omega_{pe}$ and $ 1.6 c/ \omega_{pe}$ respectively.
Solid symbols with error bars are MRX experimental 
measurements, while open squares are PIC simulation results --
both from \citet{ji}. Here $T=5$ eV and $B=0.05$ Tesla.}
\end{figure}

\section{Discussion}

Let us estimate the coronal heating flux which is produced
by our collisionless reconnection model.
One can obtain the heating flux (W m$^{-2}$) by taking
magnetic energy density $B^2/(2\mu_0)$ (J m$^{-3}$), multiplying it by a
typical scale $L = 10^5$ m and dividing by the reconnection times
which we estimated above, $t_{c'less}= \tau_A/M_{c'less}=114$ s (2 min) and $t_{sp}=
\tau_A/M_{sp}=8831$ s (0.1 days) -- assuming the reconnection time
is a good proxy for the energy release time in the solar corona.
Fixing $B=0.01$ T, in the case of collisionless reconnection, the heating flux
is obtained $3.5 \times 10^4$ W m$^{-2}$, which is more than enough
to meet the coronal heating requirement; while as expected Sweet-Parker model
produces an equivalent of 450 W m$^{-2}$. 
\citet{asch}, p. 359 quotes solar active region typical heating requirement value of
2000 W m$^{-2}$.  The obtained collisionless reconnection heating flux
is an order of magnitude larger than that. Naturally, neither all of the 
collisionless reconnection heating flux will go into dissipation
(by definition collisionless process is dissipationless), nor all 100\% of the
magnetic energy will be released by the reconnection process. Even if 1/10th of it
will ultimately end up as the heat, then the coronal heating requirement can be met.
Thus, the importance of the present model for the coronal heating seems evident.

Next, let us comment on the observational consequences of the collisionless reconnection
model in solar coronal context. The reconnection time (the flare energy release time), 
$t_{c'less}= \tau_A / M_{c'less} \propto L^2$ (because, $\tau_A = L / V_A$, and $M_{c'less}\propto 1/L$).
Thus, if we assume that magnetic field and plasma density remains the same as we
go from small to large flares (this assumption implies that for simplicity of the
argument $V_A$ in $\tau_A$ and $c/ \omega_{pe}$ in Eq.(12) stay the same),
the flare time should increase (quadratically) with flare size.
There is some ambiguity how to define actual spatial size of a flare 
(Dr. E. Kontar of University of Glasgow, 
private communication): At high energies one sees loop footpoints only;
Hence is the geometric size a distance between the footpoints or total (unknown) area?
In soft X-rays (6-10 keV) only the looptop source is observed.
X-Ray telescope (XRT) on board of Japanese space mission Hinode 
often sees the entire flaring loop. However, this mission is quite recent and no
good flare statistics exists as yet. \citet{Temmer} claim that 
the flare duration increases with increasing the flare
importance class, hence with the flare area, as
reported by a number of papers (see their Table 7).
Based on a statistical analysis of almost 50 000 soft X-ray flares observed during the period
1976-2000 \citet{Veronig} show that the
scatter plots of the flare duration, rise time
 and decay time as function of
the flare fluence (J m$^{-2}$) show strong correlations.
Whether the latter is a good proxy of the flare geometrical size is not entirely certain.
\citet{ny06}
report a statistical study of flares observed with the Soft X-Ray Telescope (SXT) on 
board of Yohkoh in the year 2000. They measured physical parameters of 77 flares.
Their Fig.~(3) plots the spatial scale $L$ against the flare timescale  and 
shows that the spatial scale $L$ tends to be larger with the increasing timescale.
Thus, in summary, our scaling seems to be in agreement with the solar flare observations.
One should be aware of the fact that in the solar flare observations
we see only post-reconnection dynamics and the
 geometric size of "particle accelerator", the location of actual 
 energy deposition site is never resolved, due to poor spatial resolution of
 the instruments.
 Some source of a concern is also the aspect ratio of the diffusion region (i.e. the current layer, 
 where particles are accelerated).
 From Eq.(14) it is clear for $M_{ c'less}=1.3 \times 10^{-4}$ the aspect ratio is
 about 7500 : 1. The question is whether such fine scale structure can survive
 in the turbulent corona. However, since current observations do not have
 enough resolution to prove or disprove existence if such elongated 
 current layers, the jury is 
 still out.
 
 On the laboratory plasma side, comparison of our model scaling with the MRX
 data (see Fig.(2)) needs a clarification as to why PIC simulation results
 (open squares) do not fit the experimental data while our simple analytical model
 does. At first sight one might expect the opposite to be true
 because PIC simulation includes all relevant physics, i.e.
 all relevant terms in the generalized Ohm's law, time dependence, etc.
 In  author's opinion the source of the discrepancy are the boundary conditions and the mass ratio mismatch.
 \citet{ji} (see their Fig.~3) either use boundary conditions commensurate to MRX (conducting 
 boundary conditions for electromagnetic fields and elastic reflection of particles
 at wall) for unrealistic mass ratio of 400, or open boundary conditions for
 a realistic mass ratio of 1836 (for Hydrogen only).
 It is no surprise that the outcomes of numerical simulations
 depend on correct (appropriate) boundary conditions used.
 There are two issues here: first, when  MRX boundary conditions are used, 
 it may well be that a better fit could have been achieved,
 if the correct mass ratio were used (note that experimental data points lie
 on the line ($8 c/ \omega_{pe}$) that is factor of  $\approx$ 5 above PIC simulation
 line $1.6 c/ \omega_{pe}$. Accidentally such is also 
 the mass ratio mismatch $1836./400=4.6$. 
 Note, however, that the 
 theoretically derived width
 according to Eq.(14) scales as $\propto 1 / (\omega_{pe}^2 r_{L,e}) \propto m_e \times 
 \omega_{ce} / v_{th,e} \propto m_e \times m_e^{-1} \times \sqrt{m_e} \propto \sqrt{m_e}$).
 Hence if our analytical treatment is correct, Eq.(14)
 still does not alleviate the problem in full, as
 $\sqrt{1836./400}=2.1$ cannot account for the  
 factor of 5 difference between the MRX data and PIC simulations. 
 But the trend is in the right direction;
 Secondly, when the correct mass ratio is used (in Ref.\cite{ji}, Fig.3) open boundary conditions are
 not a good representation of the experiment. 
 Hence the results based on open boundary conditions should be discarded.
 Thus, we conjecture that when simultaneously (a)
 the correct mass ratio and (b) boundary conditions commensurate to MRX are used, 
 than  PIC simulation results will possibly follow  our theoretical result (Eq.(14)).
 Until such simulation is performed, however, the jury in this issue is still out.

 Some clarification is necessary concerning the use of notation for $L$.
In Eq.(11) it plays a role of the {\it electron} diffusion region length; 
i.e. the region of space where the electron dynamics is decoupled from 
that of ions. Since in the solar flare observations
the geometric size of "particle accelerator", the location of actual 
 energy deposition site is never resolved, due to poor spatial 
 resolution of  the instruments, we use $L=10^5$ m, a typical length 
 scale in the corona. In context of MRX, we use the length of the 
 magnetic scale $L=0.4$ m. However, on the MHD scale (to which our 
 numerical estimates apply)
one also expects that 
at some distance downstream from the X-point, electron and ion 
dynamics becomes coupled again. This implies that the outflowing electrons
will decelerate from the electron Alfv\'en speed to ion Alfv\'en seed.
Recent PIC simulations \cite{do06,sh07} suggest that the electron 
diffusion region length can extend for much longer distances 
downstream than previously thought, and hence alleviating, 
in part, the above problem.

\section{Conclusions}

A new model of collisionless reconnection is formulated that is based on simple conservation
laws. The obtained collisionless reconnection rate, 
$M_{c'less}=( c / \omega_{pe})^2 / (r_{L,e} L)$, naturally does not depend on Lundquist number, $S$,
(because it is collisionless) and  
is much faster than
the Sweet-Parker rate, but yet somewhat slower than the Petschek rate (at least
for solar coronal plasma parameters). 
In particular, for the same set of parameters e.g.
for solar coronal plasma the rates are 
$M_{petschek}=0.04$, 
$M_{ c'less}=1.3 \times 10^{-4}$, 
$M_{sp}= 1.6 \times 10^{-6}$. 
The main implication for solar flares is that if this collisionless rate is
used, flare time can be as short as a few minutes, that is commensurate with the observations.
Note that the formulated collisionless reconnection model is general and
is not specific to any initial configuration e.g. Harris type tearing unstable current sheet,
X-point collapse or otherwise. In differ to previous results 
it relates the reconnection rate to simple, generic spatial scales
such as electron inertial length, Larmor radius and global
reconnection length. Therefore it is easily applicable to different
space or laboratory plasma situations.

\begin{acknowledgments}
Author would like to thank Prof. M. Yamada and his team at MRX
Princeton Plasma Physics Laboratory for kindly providing their experimental data set
and for useful replies to queries. Useful discussions with Dr. G. Vekstein 
and other participants of Fifths International Cambridge Workshop on Magnetic Reconnection 2008,
Bad Honnef, Germany are acknowledged.
Author acknowledges useful discussion of solar flare observational aspects with Dr. E. Kontar.
Author also would like to thank an anonymous referee who helped to improve 
the paper significantly.
This research was supported by the United Kingdom's Science and 
Technology Facilities Council (STFC).
\end{acknowledgments}


\begin{thebibliography}{25}
\expandafter\ifx\csname natexlab\endcsname\relax\def\natexlab#1{#1}\fi
\expandafter\ifx\csname bibnamefont\endcsname\relax
  \def\bibnamefont#1{#1}\fi
\expandafter\ifx\csname bibfnamefont\endcsname\relax
  \def\bibfnamefont#1{#1}\fi
\expandafter\ifx\csname citenamefont\endcsname\relax
  \def\citenamefont#1{#1}\fi
\expandafter\ifx\csname url\endcsname\relax
  \def\url#1{\texttt{#1}}\fi
\expandafter\ifx\csname urlprefix\endcsname\relax\def\urlprefix{URL }\fi
\providecommand{\bibinfo}[2]{#2}
\providecommand{\eprint}[2][]{\url{#2}}

\bibitem[{\citenamefont{{Yamada} et~al.}(2006)}]{yamada}
\bibinfo{author}{\bibfnamefont{M.}~\bibnamefont{{Yamada}}},
  \bibinfo{author}{\bibfnamefont{Y.}~\bibnamefont{{Ren}}},
  \bibinfo{author}{\bibfnamefont{H.}~\bibnamefont{{Ji}}},
  \bibinfo{author}{\bibfnamefont{J.}~\bibnamefont{{Breslau}}},
  \bibinfo{author}{\bibfnamefont{S.}~\bibnamefont{{Gerhardt}}},
  \bibinfo{author}{\bibfnamefont{R.}~\bibnamefont{{Kulsrud}}},
  \bibinfo{author}{\bibfnamefont{A.}~\bibnamefont{{Kuritsyn}}}, 
  \bibinfo{journal}{Phys. Plasmas}
  \textbf{\bibinfo{volume}{13}}, \bibinfo{pages}{052119}
  (\bibinfo{year}{2006}).

\bibitem[{\citenamefont{{Ji} et~al.}(2008)}]{ji}
\bibinfo{author}{\bibfnamefont{H.}~\bibnamefont{{Ji}}},
  \bibinfo{author}{\bibfnamefont{Y.}~\bibnamefont{{Ren}}},
  \bibinfo{author}{\bibfnamefont{M.}~\bibnamefont{{Yamada}}},
  \bibinfo{author}{\bibfnamefont{S.}~\bibnamefont{{Dorfman}}},
  \bibinfo{author}{\bibfnamefont{W.}~\bibnamefont{{Daughton}}},
  \bibnamefont{and} \bibinfo{author}{\bibfnamefont{S.~P.}
  \bibnamefont{{Gerhardt}}}, \bibinfo{journal}{Geophys. Res. Lett.}
  \textbf{\bibinfo{volume}{35}}, \bibinfo{pages}{13106} (\bibinfo{year}{2008}).

\bibitem[{\citenamefont{{Priest} and {Forbes}}(2000)}]{pf00}
\bibinfo{author}{\bibfnamefont{E.}~\bibnamefont{{Priest}}} \bibnamefont{and}
  \bibinfo{author}{\bibfnamefont{T.}~\bibnamefont{{Forbes}}},
  \emph{\bibinfo{title}{{Magnetic reconnection: MHD theory and applications}}}
  (\bibinfo{publisher}{Cambridge University Press}, \bibinfo{year}{2000}).

\bibitem[{\citenamefont{{Biskamp}}(2005)}]{biskamp}
\bibinfo{author}{\bibfnamefont{D.}~\bibnamefont{{Biskamp}}},
  \emph{\bibinfo{title}{{Magnetic reconnection in Plasmas}}}
  (\bibinfo{publisher}{Cambridge University Press}, \bibinfo{year}{2005}).

\bibitem[{\citenamefont{{Birn} and {Priest}}(2007)}]{bp07}
\bibinfo{author}{\bibfnamefont{J.}~\bibnamefont{{Birn}}} \bibnamefont{and}
  \bibinfo{author}{\bibfnamefont{E.~R.} \bibnamefont{{Priest}}},
  \emph{\bibinfo{title}{{Reconnection of magnetic fields: magnetohydrodynamics
  and collisionless theory and observations}}} (\bibinfo{publisher}{Cambridge
  University Press}, \bibinfo{year}{2007}).

\bibitem[{\citenamefont{{Innes} et~al.}(1997)}]{1997Natur.386..811I}
\bibinfo{author}{\bibfnamefont{D.~E.} \bibnamefont{{Innes}}},
  \bibinfo{author}{\bibfnamefont{B.}~\bibnamefont{{Inhester}}},
  \bibinfo{author}{\bibfnamefont{W.~I.} \bibnamefont{{Axford}}},
   \bibinfo{journal}{Nature} \textbf{\bibinfo{volume}{386}},
  \bibinfo{pages}{811} (\bibinfo{year}{1997}).

\bibitem[{\citenamefont{{Woods}}(2008)}]{2008AdSpR..42..895W}
\bibinfo{author}{\bibfnamefont{T.~N.} \bibnamefont{{Woods}}},
  \bibinfo{journal}{Adv. Sp. Res.} \textbf{\bibinfo{volume}{42}},
  \bibinfo{pages}{895} (\bibinfo{year}{2008}).

\bibitem[{\citenamefont{{Schmitt} et~al.}(2008)}]{2008A&A...481..799S}
\bibinfo{author}{\bibfnamefont{J.~H.~M.~M.} \bibnamefont{{Schmitt}}},
  \bibinfo{author}{\bibfnamefont{F.}~\bibnamefont{{Reale}}},
  \bibinfo{author}{\bibfnamefont{C.}~\bibnamefont{{Liefke}}},
  \bibinfo{author}{\bibfnamefont{U.}~\bibnamefont{{Wolter}}},
  \bibinfo{author}{\bibfnamefont{B.}~\bibnamefont{{Fuhrmeister}}},
  \bibinfo{author}{\bibfnamefont{A.}~\bibnamefont{{Reiners}}},
  \bibnamefont{and} \bibinfo{author}{\bibfnamefont{G.}~\bibnamefont{{Peres}}},
  \bibinfo{journal}{Astron.  Astrophys.} \textbf{\bibinfo{volume}{481}},
  \bibinfo{pages}{799} (\bibinfo{year}{2008}).

\bibitem[{\citenamefont{{Cassak} et~al.}(2008)}]{2008ApJ...676L..69C}
\bibinfo{author}{\bibfnamefont{P.~A.} \bibnamefont{{Cassak}}},
  \bibinfo{author}{\bibfnamefont{D.~J.} \bibnamefont{{Mullan}}},
  \bibnamefont{and} \bibinfo{author}{\bibfnamefont{M.~A.}
  \bibnamefont{{Shay}}}, \bibinfo{journal}{Astrophys. J.}
  \textbf{\bibinfo{volume}{676}}, \bibinfo{pages}{L69} (\bibinfo{year}{2008}).

\bibitem[{\citenamefont{{Runov} et~al.}(2008)}]{2008JGRA..11307S27R}
\bibinfo{author}{\bibfnamefont{A.}~\bibnamefont{{Runov}}},
  \bibinfo{author}{\bibfnamefont{W.}~\bibnamefont{{Baumjohann}}},
  \bibinfo{author}{\bibfnamefont{R.}~\bibnamefont{{Nakamura}}},
  \bibinfo{author}{\bibfnamefont{V.~A.} \bibnamefont{{Sergeev}}},
  \bibinfo{author}{\bibfnamefont{O.}~\bibnamefont{{Amm}}},
  \bibinfo{author}{\bibfnamefont{H.}~\bibnamefont{{Frey}}},
  \bibinfo{author}{\bibfnamefont{I.}~\bibnamefont{{Alexeev}}},
  \bibinfo{author}{\bibfnamefont{A.~N.} \bibnamefont{{Fazakerley}}},
  \bibinfo{author}{\bibfnamefont{C.~J.} \bibnamefont{{Owen}}},
  \bibinfo{author}{\bibfnamefont{E.}~\bibnamefont{{Lucek}}},
   \bibinfo{journal}{J. Geophys. Res.} \textbf{\bibinfo{volume}{113}}, \bibinfo{pages}{7}
  (\bibinfo{year}{2008}).

\bibitem[{\citenamefont{{Sweet}}(1958)}]{sweet}
\bibinfo{author}{\bibfnamefont{P.~A.} \bibnamefont{{Sweet}}},
  \bibinfo{journal}{Nuovo \ Cim. \ Suppl.} \textbf{\bibinfo{volume}{8}},
  \bibinfo{pages}{188} (\bibinfo{year}{1958}).

\bibitem[{\citenamefont{{Parker}}(1963)}]{parker}
\bibinfo{author}{\bibfnamefont{E.~N.} \bibnamefont{{Parker}}},
  \bibinfo{journal}{Astrophys. \ J. \ Suppl. \ Ser.}
  \textbf{\bibinfo{volume}{8}}, \bibinfo{pages}{177} (\bibinfo{year}{1963}).

\bibitem[{\citenamefont{{Petschek}}(1964)}]{petschek}
\bibinfo{author}{\bibfnamefont{H.~E.} \bibnamefont{{Petschek}}}, in
  \emph{\bibinfo{booktitle}{In AAS/NASA Symposium on the Physics of Solar
  Flares}}, edited by \bibinfo{editor}{\bibfnamefont{W.~N.}
  \bibnamefont{{Hess}}} (\bibinfo{publisher}{NASA Press, Washington, DC},
  \bibinfo{year}{1964}), pp. \bibinfo{pages}{425--437}.

\bibitem[{\citenamefont{{Hesse} et~al.}(1999)}]{hesse99}
\bibinfo{author}{\bibfnamefont{M.}~\bibnamefont{{Hesse}}},
  \bibinfo{author}{\bibfnamefont{K.}~\bibnamefont{{Schindler}}},
  \bibinfo{author}{\bibfnamefont{J.}~\bibnamefont{{Birn}}}, \bibnamefont{and}
  \bibinfo{author}{\bibfnamefont{M.}~\bibnamefont{{Kuznetsova}}},
  \bibinfo{journal}{Phys. Plasmas} \textbf{\bibinfo{volume}{6}},
  \bibinfo{pages}{1781} (\bibinfo{year}{1999}).

\bibitem[{\citenamefont{{Birn} et~al.}(2001)}]{birn01}
\bibinfo{author}{\bibfnamefont{J.}~\bibnamefont{{Birn}}},
  \bibinfo{author}{\bibfnamefont{J.~F.} \bibnamefont{{Drake}}},
  \bibinfo{author}{\bibfnamefont{M.~A.} \bibnamefont{{Shay}}},
  \bibinfo{author}{\bibfnamefont{B.~N.} \bibnamefont{{Rogers}}},
  \bibinfo{author}{\bibfnamefont{R.~E.} \bibnamefont{{Denton}}},
  \bibinfo{author}{\bibfnamefont{M.}~\bibnamefont{{Hesse}}},
  \bibinfo{author}{\bibfnamefont{M.}~\bibnamefont{{Kuznetsova}}},
  \bibinfo{author}{\bibfnamefont{Z.~W.} \bibnamefont{{Ma}}},
  \bibinfo{author}{\bibfnamefont{A.}~\bibnamefont{{Bhattacharjee}}},
  \bibinfo{author}{\bibfnamefont{A.}~\bibnamefont{{Otto}}},
  \bibinfo{journal}{J. \ Geophys. \ Res.}
  \textbf{\bibinfo{volume}{106}}, \bibinfo{pages}{3715} (\bibinfo{year}{2001}).

\bibitem[{\citenamefont{{Pritchett}}(2001)}]{pritchett01}
\bibinfo{author}{\bibfnamefont{P.~L.} \bibnamefont{{Pritchett}}},
  \bibinfo{journal}{J. \ Geophys. \ Res.} \textbf{\bibinfo{volume}{106}},
  \bibinfo{pages}{3783} (\bibinfo{year}{2001}).

\bibitem[{\citenamefont{{Tsiklauri} and {Haruki}}(2007)}]{th07}
\bibinfo{author}{\bibfnamefont{D.}~\bibnamefont{{Tsiklauri}}} \bibnamefont{and}
  \bibinfo{author}{\bibfnamefont{T.}~\bibnamefont{{Haruki}}},
  \bibinfo{journal}{Phys. Plasmas} \textbf{\bibinfo{volume}{14}},
  \bibinfo{pages}{112905} (\bibinfo{year}{2007}).

\bibitem[{\citenamefont{{Tsiklauri} and {Haruki}}(2008)}]{th08}
\bibinfo{author}{\bibfnamefont{D.}~\bibnamefont{{Tsiklauri}}} \bibnamefont{and}
  \bibinfo{author}{\bibfnamefont{T.}~\bibnamefont{{Haruki}}},
  \bibinfo{journal}{{\it Physics of collisionless reconnection in a stressed
  X-point collapse}, Phys. Plasmas (accepted, inpress)}  (\bibinfo{year}{2008}).

\bibitem[{\citenamefont{{Kuznetsova} et~al.}(1998)}]{kuz98}
\bibinfo{author}{\bibfnamefont{M.~M.} \bibnamefont{{Kuznetsova}}},
  \bibinfo{author}{\bibfnamefont{M.}~\bibnamefont{{Hesse}}}, \bibnamefont{and}
  \bibinfo{author}{\bibfnamefont{D.}~\bibnamefont{{Winske}}},
  \bibinfo{journal}{J. \ Geophys. \ Res. } \textbf{\bibinfo{volume}{103}}, \bibinfo{pages}{199}
  (\bibinfo{year}{1998}).

\bibitem[{\citenamefont{{Birn} and {Hesse}}(2007)}]{birn07}
\bibinfo{author}{\bibfnamefont{J.}~\bibnamefont{{Birn}}} \bibnamefont{and}
  \bibinfo{author}{\bibfnamefont{M.}~\bibnamefont{{Hesse}}},
  \bibinfo{journal}{Phys. Plasmas} \textbf{\bibinfo{volume}{14}},
  \bibinfo{pages}{082306} (\bibinfo{year}{2007}).

\bibitem[{\citenamefont{{Klimas} et~al.}(2008)}]{klimas08}
\bibinfo{author}{\bibfnamefont{A.}~\bibnamefont{{Klimas}}},
  \bibinfo{author}{\bibfnamefont{M.}~\bibnamefont{{Hesse}}}, \bibnamefont{and}
  \bibinfo{author}{\bibfnamefont{S.}~\bibnamefont{{Zenitani}}},
  \bibinfo{journal}{Phys. Plasmas} \textbf{\bibinfo{volume}{15}},
  \bibinfo{pages}{082102} (\bibinfo{year}{2008}).

\bibitem[{\citenamefont{{Vekstein} and {Priest}}(1995)}]{vp95}
\bibinfo{author}{\bibfnamefont{G.~E.} \bibnamefont{{Vekstein}}}
  \bibnamefont{and} \bibinfo{author}{\bibfnamefont{E.~R.}
  \bibnamefont{{Priest}}}, \bibinfo{journal}{Phys. Plasmas}
  \textbf{\bibinfo{volume}{2}}, \bibinfo{pages}{3169} (\bibinfo{year}{1995}).

\bibitem[{\citenamefont{{Golub} and {Pasachoff}}(1997)}]{golub}
\bibinfo{author}{\bibfnamefont{L.}~\bibnamefont{{Golub}}} \bibnamefont{and}
  \bibinfo{author}{\bibfnamefont{J.~M.} \bibnamefont{{Pasachoff}}},
  \emph{\bibinfo{title}{{The Solar Corona}}} (\bibinfo{publisher}{Cambridge
  University Press, UK}, \bibinfo{year}{1997}).

\bibitem[{\citenamefont{{Horiuchi} and {Sato}}(1997)}]{hs97}
\bibinfo{author}{\bibfnamefont{R.}~\bibnamefont{{Horiuchi}}} \bibnamefont{and}
  \bibinfo{author}{\bibfnamefont{T.}~\bibnamefont{{Sato}}},
  \bibinfo{journal}{Phys. Plasmas} \textbf{\bibinfo{volume}{4}},
  \bibinfo{pages}{277} (\bibinfo{year}{1997}).

\bibitem[{\citenamefont{{Breslau} and {Jardin}}(2003)}]{breslau}
\bibinfo{author}{\bibfnamefont{J.~A.} \bibnamefont{{Breslau}}}
  \bibnamefont{and} \bibinfo{author}{\bibfnamefont{S.~C.}
  \bibnamefont{{Jardin}}}, \bibinfo{journal}{Phys. Plasmas}
  \textbf{\bibinfo{volume}{10}}, \bibinfo{pages}{1291} (\bibinfo{year}{2003}).
  
  
\bibitem[{\citenamefont{{Aschwanden}}(2006)}]{asch}
\bibinfo{author}{\bibfnamefont{M.~J.}~\bibnamefont{{Aschwanden}}},
  \emph{\bibinfo{title}{{Physics of the Solar Corona an Introduction}}} 
  (\bibinfo{publisher}{Paxis Publishing Ltd, Chichester, UK}, \bibinfo{year}{2006}).
  
\bibitem[{\citenamefont{{Temmer} et~al.}(2001)}]{Temmer}
\bibinfo{author}{\bibfnamefont{M.}~\bibnamefont{{Temmer}}},
  \bibinfo{author}{\bibfnamefont{A.}~\bibnamefont{{Veronig}}}, 
  \bibinfo{author}{\bibfnamefont{A.}~\bibnamefont{{Hanslmeier}}},
   \bibinfo{author}{\bibfnamefont{W.}~\bibnamefont{{Otruba}}}  \bibnamefont{and}
   \bibinfo{author}{\bibfnamefont{M.}~\bibnamefont{{Messerotti}}},
  \bibinfo{journal}{Astron.  Astrophys.} \textbf{\bibinfo{volume}{375}},
  \bibinfo{pages}{1049} (\bibinfo{year}{2001}). 
   
\bibitem[{\citenamefont{{Veronig} et~al.}(2002)}]{Veronig}
\bibinfo{author}{\bibfnamefont{A.}~\bibnamefont{{Veronig}}},
  \bibinfo{author}{\bibfnamefont{M.}~\bibnamefont{{Temmer}}}, 
  \bibinfo{author}{\bibfnamefont{A.}~\bibnamefont{{Hanslmeier}}},
   \bibinfo{author}{\bibfnamefont{W.}~\bibnamefont{{Otruba}}}  \bibnamefont{and}
   \bibinfo{author}{\bibfnamefont{M.}~\bibnamefont{{Messerotti}}},
  \bibinfo{journal}{Astron.  Astrophys.} \textbf{\bibinfo{volume}{382}},
  \bibinfo{pages}{1070} (\bibinfo{year}{2002}).   

  
  \bibitem[{\citenamefont{{Nagashima} and {Yokoyama}}(2006)}]{ny06}
\bibinfo{author}{\bibfnamefont{K.}~\bibnamefont{{Nagashima}}} \bibnamefont{and}
  \bibinfo{author}{\bibfnamefont{T.}~\bibnamefont{{Yokoyama}}},
  \bibinfo{journal}{Astrophys. J.} \textbf{\bibinfo{volume}{647}},
  \bibinfo{pages}{654} (\bibinfo{year}{2006}).
  
 
 \bibitem[{\citenamefont{{Daughton} et~al.}(2006)}]{do06}
\bibinfo{author}{\bibfnamefont{W.}~\bibnamefont{{Daughton}}},
  \bibinfo{author}{\bibfnamefont{J.}~\bibnamefont{{Scudder}}}, \bibnamefont{and}
  \bibinfo{author}{\bibfnamefont{H.}~\bibnamefont{{Karimabadi}}},
   \bibinfo{journal}{Phys. Plasmas} \textbf{\bibinfo{volume}{13}},
  \bibinfo{pages}{072101} (\bibinfo{year}{2006}).
  
  
  \bibitem[{\citenamefont{{Shay} et~al.}(2007)}]{sh07}
\bibinfo{author}{\bibfnamefont{M.~A.}~\bibnamefont{{Shay}}},
  \bibinfo{author}{\bibfnamefont{J.~F.}~\bibnamefont{{Drake}}}, \bibnamefont{and}
  \bibinfo{author}{\bibfnamefont{M.}~\bibnamefont{{Swisdak}}},
   \bibinfo{journal}{Phys. Rev. Lett.} \textbf{\bibinfo{volume}{99}},
  \bibinfo{pages}{155002} (\bibinfo{year}{2007}).
 
 \end{thebibliography}
\end{document}